\newcommand\apjcls{1}
\newcommand\aastexcls{2}
\newcommand\othercls{3}

\newcommand\papercls{\aastexcls}
\documentclass[tighten, times, twocolumn]{aastex7}  

\usepackage{lineno}
\if\papercls \apjcls
\usepackage{apjfonts}
\else\if\papercls \othercls
\usepackage{epsfig}
\usepackage{margin}
\usepackage{times}
\fi\fi
\usepackage{ifthen}
\usepackage{natbib}
\usepackage{amssymb, amsmath}
\usepackage{appendix}
\usepackage{etoolbox}
\usepackage[T1]{fontenc}
\usepackage{paralist}
\everymath{\displaystyle}

\if\papercls \apjcls
\newcommand\aas{\ref@jnl{AAS Meeting Abstracts}}
\newcommand\dps{\ref@jnl{AAS/DPS Meeting Abstracts}}
\newcommand\maps{\ref@jnl{MAPS}}
\else\if\papercls \othercls
\usepackage{astjnlabbrev-jh}
\fi\fi

\if\papercls \aastexcls
\hypersetup{citecolor=blue, 
            linkcolor=blue, 
            menucolor=blue, 
            urlcolor=blue}  
\else
\usepackage[
bookmarks=true,           
bookmarksnumbered=true,   
colorlinks=true,          
citecolor=blue,           
linkcolor=blue,           
menucolor=blue,           
urlcolor=blue,            
linkbordercolor={0 0 1},  
pdfborder={0 0 1},
frenchlinks=true]{hyperref}
\fi

\if\papercls \othercls

\else

\fi

\providecommand{\adsurl}[1]{\href{#1}{ADS}}

\makeatletter
\patchcmd{\NAT@citex}
  {\@citea\NAT@hyper@{%
     \NAT@nmfmt{\NAT@nm}%
     \hyper@natlinkbreak{\NAT@aysep\NAT@spacechar}{\@citeb\@extra@b@citeb}%
     \NAT@date}}
  {\@citea\NAT@nmfmt{\NAT@nm}%
   \NAT@aysep\NAT@spacechar\NAT@hyper@{\NAT@date}}{}{}

\patchcmd{\NAT@citex}
  {\@citea\NAT@hyper@{%
     \NAT@nmfmt{\NAT@nm}%
     \hyper@natlinkbreak{\NAT@spacechar\NAT@@open\if*#1*\else#1\NAT@spacechar\fi}%
       {\@citeb\@extra@b@citeb}%
     \NAT@date}}
  {\@citea\NAT@nmfmt{\NAT@nm}%
   \NAT@spacechar\NAT@@open\if*#1*\else#1\NAT@spacechar\fi\NAT@hyper@{\NAT@date}}
  {}{}
\makeatother

\makeatletter
\DeclareRobustCommand{\lowcase}[1]{\@lowcase#1\@nil}
\def\@lowcase#1\@nil{\if\relax#1\relax\else\MakeLowercase{#1}\fi}
\makeatother

\DeclareSymbolFont{UPM}{U}{eur}{m}{n}
\DeclareMathSymbol{\umu}{0}{UPM}{"16}
\let\oldumu=\umu
\renewcommand\umu{\ifmmode\oldumu\else\math{\oldumu}\fi}

\if\papercls \othercls

\else

\fi

\let\oldsim=\sim
\renewcommand\sim{\ifmmode\oldsim\else\math{\oldsim}\fi}
\let\oldpm=\pm
\renewcommand\pm{\ifmmode\oldpm\else\math{\oldpm}\fi}
\newcommand\by{\ifmmode\times\else\math{\times}\fi}

\newbox{\wdbox}
\renewcommand\c{\setbox\wdbox=\hbox{,}\hspace{\wd\wdbox}}
\renewcommand\i{\setbox\wdbox=\hbox{i}\hspace{\wd\wdbox}}
\newcommand\n{\hspace{0.5em}}

\newcount\timect
\newcount\hourct
\newcount\minct
\newcommand\now{\timect=\time \divide\timect by 60
         \hourct=\timect \multiply\hourct by 60
         \minct=\time \advance\minct by -\hourct
         \number\timect:\ifnum \minct < 10 0\fi\number\minct}

\catcode`@=11

\newcommand\comment[1]{}

\newcommand\commenton{\catcode`\%=14}

\renewcommand\math[1]{$#1$}
\newcommand\mathshifton{\catcode`\$=3}

\let\atab=&
\newcommand\atabon{\catcode`\&=4}

\let\oldmsp=\sp
\let\oldmsb=\sb
\def\sp#1{\ifmmode
           \oldmsp{#1}%
         \else\strut\raise.85ex\hbox{\scriptsize #1}\fi}
\def\sb#1{\ifmmode
           \oldmsb{#1}%
         \else\strut\raise-.54ex\hbox{\scriptsize #1}\fi}
\newbox\@sp
\newbox\@sb
\def\sbp#1#2{\ifmmode%
           \oldmsb{#1}\oldmsp{#2}%
         \else
           \setbox\@sb=\hbox{\sb{#1}}%
           \setbox\@sp=\hbox{\sp{#2}}%
           \rlap{\copy\@sb}\copy\@sp
           \ifdim \wd\@sb >\wd\@sp
             \hskip -\wd\@sp \hskip \wd\@sb
           \fi
        \fi}
\def\msp#1{\ifmmode
           \oldmsp{#1}
         \else \math{\oldmsp{#1}}\fi}
\def\msb#1{\ifmmode
           \oldmsb{#1}
         \else \math{\oldmsb{#1}}\fi}

\def\supon{\catcode`\^=7}

\def\subon{\catcode`\_=8}

\def\supsubon{\supon \subon}

\newcommand\actcharon{\catcode`\~=13}

\newcommand\paramon{\catcode`\#=6}

\comment{And now to turn us totally on and off...}

\newcommand\reservedcharson{ \commenton  \mathshifton  \atabon  \supsubon 
                             \actcharon  \paramon}

\catcode`@=12
\reservedcharson

\if\papercls \apjcls

\else

\fi

\newcommand\chisq{\ifmmode{\chi\sp{2}}\else\math{\chi\sp{2}}\fi}
\newcommand\redchisq{\ifmmode{ \chi\sp{2}\sb{\rm red}}
                    \else\math{\chi\sp{2}\sb{\rm red}}\fi}
\newcommand\Teq{\ifmmode{T\sb{\rm eq}}\else$T$\sb{eq}\fi}
\newcommand\mjup{\ifmmode{M\sb{\rm Jup}}\else$M$\sb{Jup}\fi}
\newcommand\rjup{\ifmmode{R\sb{\rm Jup}}\else$R$\sb{Jup}\fi}
\newcommand\msun{\ifmmode{M\sb{\odot}}\else$M\sb{\odot}$\fi}
\newcommand\rsun{\ifmmode{R\sb{\odot}}\else$R\sb{\odot}$\fi}
\newcommand\mearth{\ifmmode{M\sb{\oplus}}\else$M\sb{\oplus}$\fi}
\newcommand\rearth{\ifmmode{R\sb{\oplus}}\else$R\sb{\oplus}$\fi}

\shorttitle{Multi-Velocity Ultra-Fast Outflows in IRAS~05189$-$2524}
\shortauthors{Noda {\em et al.}}

\begin{document}

\title{Discovery of Powerful Multi-Velocity Ultra-Fast Outflows in the Starburst Merger Galaxy IRAS~05189$-$2524 with XRISM}

\author{Hirofumi Noda}
\affiliation{Astronomical Institute, Tohoku University, 6-3 Aramakiazaaoba, Aoba-ku, Sendai, Miyagi 980-8578, Japan}
\email[show]{hirofumi.noda@astr.tohoku.ac.jp}  

\author{Satoshi Yamada}
\affiliation{Frontier Research Institute for Interdisciplinary Sciences, Tohoku University, 6-3 Aramakiazaaoba, Aoba-ku, Sendai, Miyagi 980-8578, Japan}
\affiliation{Astronomical Institute, Tohoku University, 6-3 Aramakiazaaoba, Aoba-ku, Sendai, Miyagi 980-8578, Japan}
\email[]{}  

\author{Shoji Ogawa}
\affiliation{Institute of Space and Astronautical Science, Japan Aerospace Exploration Agency, 3-1-1 Yoshino-dai, Chuo-ku, Sagamihara, Kanagawa 252-5210, Japan}
\email[]{}  

\author{Kouichi Hagino}
\affiliation{Department of Physics, University of Tokyo, 7-3-1 Hongo, Bunkyo-ku, Tokyo 113-0033, Japan}
\email[]{}  

\author{Ehud Behar}
\affiliation{Department of Physics, Technion, Haifa 3200003, Israel}
\email[]{}  

\author{Omer Reich}
\affiliation{Department of Physics, Technion, Haifa 3200003, Israel}
\email[]{}  

\author{Anna Ogorzalek}
\affiliation{Department of Astronomy, University of Maryland, MD 20742, USA}
\affiliation{NASA/Goddard Space Flight Center, MD 20771, USA}
\affiliation{Center for Research and Exploration in Space Science and Technology, NASA/GSFC (CRESST II), MD 20771, USA}
\email[]{}  

\author{Laura Brenneman}
\affiliation{Center for Astrophysics -- Harvard-Smithsonian, MA 02138, USA}
\email[]{}  

\author{Yuichi Terashima}
\affiliation{Department of Physics, Ehime University, Ehime 790-8577, Japan}
\email[]{}  

\author{Misaki Mizumoto}
\affiliation{Science Research Education Unit, University of Teacher Education Fukuoka, Fukuoka 811-4192, Japan}
\email[]{}  

\author{Francesco Tombesi}
\affiliation{Department of Physics, University of Rome Tor Vergata, Rome, Italy}
\email[]{}  

\author{Pierpaolo Cond\`{o}}
\affiliation{Department of Physics, University of Rome Tor Vergata, Rome, Italy}
\email[]{}  

\author{Alfredo Luminari}
\affiliation{INAF -- Istituto di Astrofisica e Planetologia Spaziali, Rome, Italy}
\affiliation{INAF -- Osservatorio Astronomico di Roma, Monte Porzio Catone, Italy}
\email[]{}  

\author{Atsushi Tanimoto}
\affiliation{Graduate School of Science and Engineering, Kagoshima University, Kagoshima 890-8580, Japan}
\email[]{}  

\author{Megan E. Eckart}
\affiliation{Lawrence Livermore National Laboratory, CA 94550, USA}
\email[]{}  

\author{Erin Kara}
\affiliation{Kavli Institute for Astrophysics and Space Research, Massachusetts Institute of Technology, MA 02139, USA}
\email[]{}  

\author{Takashi Okajima}
\affiliation{NASA/Goddard Space Flight Center, MD 20771, USA}
\email[]{}  

\author{Yoshihiro Ueda}
\affiliation{Department of Astronomy, Kyoto University, Kyoto 606-8502, Japan}
\email[]{}  

\author{Yuki Aiso}
\affiliation{Department of Astronomy, Kyoto University, Kyoto 606-8502, Japan}
\email[]{}  

\author{Makoto Tashiro}
\affiliation{Department of Physics, Saitama University, Saitama 338-8570, Japan}
\affiliation{Institute of Space and Astronautical Science, Japan Aerospace Exploration Agency, 3-1-1 Yoshino-dai, Chuo-ku, Sagamihara, Kanagawa 252-5210, Japan}
\email[]{}  



\begin{abstract}
We observed the X-ray-bright ultra-luminous infrared galaxy, IRAS~05189$-$2524, with XRISM during its performance verification phase. 
The unprecedented energy resolution of the onboard X-ray microcalorimeter revealed complex spectral features at $\sim7$--9~keV, which can be interpreted as blueshifted Fe~XXV/XXVI absorption lines with various velocity dispersions, originating from ultra-fast outflow (UFO) components with multiple bulk velocities of $\sim0.076c$, $\sim0.101c$, and $\sim0.143c$. 
In addition, a broad Fe-K emission line was detected around $\sim7$~keV, forming a P~Cygni profile together with the absorption lines. 
The onboard X-ray CCD camera revealed a 0.4--12~keV broadband spectrum characterized by a neutrally absorbed power-law continuum with a photon index of $\sim$2.3, and intrinsic flare-like variability on timescales of $\sim10$~ksec, both of which are likely associated with near-Eddington accretion. 
We also found potential variability of the UFO parameters on a timescale of $\sim140$~ksec.
Using these properties, we propose new constraints on the outflow structure and suggest the presence of multiple outflowing regions on scales of about tens to a hundred Schwarzschild radii, located within roughly two thousand Schwarzschild radii.
Since both the estimated momentum and energy outflow rates of the UFOs exceed those of galactic molecular outflows, our results indicate that powerful, multi-velocity UFOs are already well developed during a short-lived evolutionary phase following a major galaxy merger, characterized by intense starburst activity and likely preceding the quasar phase. 
This system is expected to evolve into a quasar, sustaining strong UFO activity and suppressing star formation in the host galaxy.
\end{abstract}
\keywords{galaxies: active -- galaxies: individual (IRAS 05189$-$2524) -- galaxies: quasars -- X-rays: galaxies}

\section{Introduction}

The co-evolution of galaxies and their central supermassive black holes (SMBHs) is a fundamental question in modern astronomy and astrophysics (e.g., \citealt{Kormendy2013}).
Galaxy mergers are thought to play a particularly important role in this process by inducing gas inflow toward the nuclear region through gravitational torques, thereby triggering intense starburst activity and the emergence of quasars (\citealt{Hopkins2006}; \citealt{Hopkins2008}).
The resulting active galactic nucleus (AGN) feedback is believed to be a key factor in shaping the subsequent evolution of the host galaxy.
However, the effects of AGN feedback on surrounding starburst activity after the emergence of a quasar remain under much debate.

One promising feedback mechanism involves the launching of powerful disk winds, in addition to the intense X-ray and ultraviolet radiation emitted by the AGN.
These winds can interact with the surrounding interstellar medium, potentially suppressing star formation (e.g., \citealt{Fabian2012}; \citealt{Zubovas2012}). 
Among disk winds from AGNs, the most powerful cases are known as ultra-fast outflows (UFOs), which are launched at velocities of a few tens of percent of the light velocity $c$. 
Representative examples include type I quasars such as PDS~456 (e.g., \citealt{Reeves2003}) and PG~1211+143 (e.g., \citealt{Pounds2003}), where blueshifted absorption lines of Fe~XXV and Fe~XXVI have been detected at velocities of $\sim 0.1$--$0.3c$.
However, the launching mechanisms of fast disk winds, the amount of mass, momentum, and energy they carry, the evolutionary stage toward the quasar phase in which they become prominent, and their role in AGN feedback remain poorly understood, constituting one of the most important open questions in AGN research.

A system in which a highly accreting AGN capable of launching a disk wind coexists with an intense starburst provides an ideal laboratory for investigating AGN feedback, as it may capture a transitional phase preceding the emergence of a quasar, during which the feedback is actively influencing starburst activity.
A key class of targets that may satisfy these conditions is X-ray-bright ultra-luminous infrared galaxies (ULIRGs), as they are known to host both vigorous starburst activity and a luminous AGN (e.g., \citealt{Iwasawa2011}; \citealt{Koss2013}; \citealt{Yamada2021}).
IRAS~05189$-$2524 is one of the X-ray brightest ULIRGs and is a late-stage merger at a redshift of $z = 0.0426$ \citep{Veilleux2002}.
Its luminous X-ray emission suggests that a fully developed quasar has already emerged, while its quite high infrared luminosity ($\log L_{\rm IR}/L_{\odot}=12.16$; \citealt{U2012}) indicates that intense starburst activity is still ongoing.
Near-infrared spectroscopy reveals a broad P$\alpha$ emission line \citep{Severgnini2001}, whereas the optical classification places the source close to type 2.
Although the mass of the central SMBH has been estimated using the $M$--$\sigma$ relation, the constraint remains weak with reported values ranging from  $M_{\rm BH} = 2.5 \times 10^7~M_{\odot}$ \citep{Koss2022} to $4.2 \times 10^8~M_{\odot}$  \citep{Xu2017}.

IRAS~05189$-$2524 has been extensively studied in the X-ray band.
It was observed by ASCA in 1995 and BeppoSAX in 1999 \citep{Severgnini2001} and later by XMM-Newton and Chandra in the early 2000s \citep{Teng2009}.
The soft X-ray emission below $\sim 1.5$~keV is dominated by either thin-thermal plasma emission from the host galaxy or scattered continuum from the narrow-line region in the AGN.
In contrast, at energies above $\sim 1.5$~keV, an intrinsic power-law continuum from the X-ray corona near the SMBH is observed, modified by absorption with a column density of $N_{\rm H} \sim 10^{23}$~cm$^{-2}$.
Between the 1990s and early 2000s, the soft X-ray flux remained almost unchanged, and it was classified as a Compton-thin AGN.
However, in 2006, observations with Suzaku revealed the appearance of an absorber with $N_{\rm H} > 10^{24}$~cm$^{-2}$ along the line of sight, temporarily transforming the AGN into a Compton-thick state (e.g., \citealt{Teng2009}; \citealt{Yamada2021}).
In 2013, simultaneous observations with XMM-Newton and NuSTAR confirmed that the source had returned to the Compton-thin phase \citep{Xu2017}.

Multiple studies have reported on X-ray emission and absorption lines in IRAS~05189$-$2524. 
\cite{Teng2009} detected a neutral Fe-K$\alpha$ emission line at 6.4~keV during the Compton-thick state observed with Suzaku. 
Later, \cite{Teng2015} reported an additional broad emission line around 6.8~keV based on observations with XMM-Newton and NuSTAR. 
\cite{Xu2017} suggested that this broad feature might be a relativistically blurred Fe-K$\alpha$ profile.
Furthermore, \cite{Smith2019} reported an absorption line at $\sim 7.5$~keV in XMM-Newton spectra, which was apparently attributed to an UFO with a velocity of $\sim 0.1c$. 
This feature was interpreted as an Fe XXVI absorption line; however, due to the limited energy resolution, its width remained unconstrained, with an upper limit of $\sigma < 240$~eV.

On September 7, 2023 (JST), the X-Ray Imaging and Spectroscopy Mission (XRISM; \citealt{Tashiro2025}) was successfully launched.
With its onboard X-ray microcalorimeter, XRISM has enabled high-resolution X-ray spectroscopy in orbit, allowing observations of a wide range of X-ray sources (e.g., \citealt{XRISM_N132D}; \citealt{XRISM_CygX3}).
During its Performance Verification (PV) phase, XRISM observed IRAS~05189$-$2524 and detected particularly remarkable spectral features. 
The high-resolution spectrum of IRAS~05189$-$2524 revealed that an absorption feature previously seen as a single broad line in lower-resolution data actually consists of multiple, narrower lines. 
This indicates the presence of UFOs with multiple velocity components.
A similar result has been reported for a type-1 quasar PDS~456, as described in \citet{XRISM2025}.
In this paper, we present the results from the XRISM observation of IRAS~05189$-$2524 and discuss their interpretations.
We focus on the significance of detecting powerful UFOs in ULIRGs, the physical structure of the outflows, including the development of velocity dispersion, and their mass, momentum, and energy outflow rates in comparison with those of previously reported molecular outflows in the host galaxy.

This paper is organized as follows.
In Section 2, we describe the XRISM observation of IRAS~05189$-$2524, along with the data reduction process for the X-ray microcalorimeter of Resolve and the X-ray CCD camera of Xtend.
In Section 3, we present the spectral and timing analyses based on the data obtained from Resolve and Xtend.
In Section 4, we summarize our results and discuss the significance of the detected UFOs in the ULIRG, their structural characteristics, and a comparison of their mass, momentum, and energy outflow rates with those of galactic molecular outflows.
We adopt  cosmological parameters of $H_0 = 73$~km~s$^{-1}$~Mpc$^{-1}$, $\Omega_{\Lambda} = 0.73$ and $\Omega_{\rm m} = 0.27$ throughout the present paper. 
Errors quoted in this paper refer to 1$\sigma$ errors in figures and $90$\% errors in tables unless noted otherwise. 

\section{Observation and Data Reduction}
\label{two}

\begin{figure*}[t]
\epsscale{1.1}
\plotone{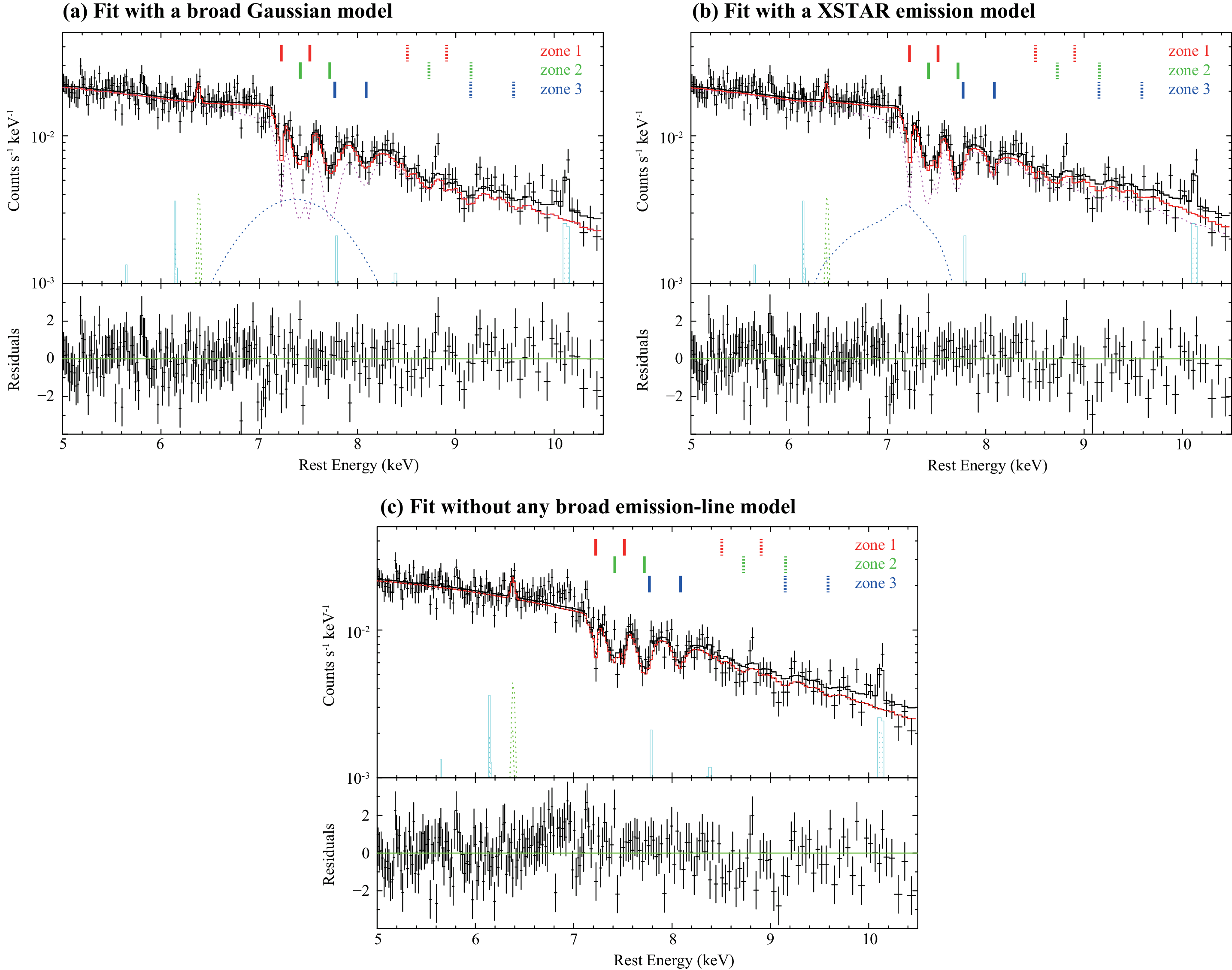}
\caption{The 5.0--10.5~keV Resolve spectrum and residuals in the fit with the model including the three-zone UFO components, shown in the rest frame. Panels (a) and (b) include a broad Gaussian and XSTAR emission model, respectively, while panel (c) shows the fit without any broad emission-line model.
Red, purple, blue, green, and cyan show the total source spectrum, power-law continuum absorbed by UFOs, broad Gaussian emission line, narrow Fe-K$\alpha$ emission line, and the NXB spectrum, respectively. Vertical solid and dotted lines indicate the approximate positions of the Fe~XXV/XXVI K$\alpha$ and K$\beta$ absorption features, respectively, in Zones 1 (red), 2 (green), and 3 (blue). }
\label{resolve_spec}
\end{figure*}

\renewcommand{\arraystretch}{1}
\begin{table*}
 \caption{Parameters of the UFO components (\texttt{XSTAR$_{\rm UFO}$}) and the statistics obtained in the fit to the 1.7--12.0~keV Resolve spectrum. }
 \label{tab1}
 \begin{center}
 \begin{tabular}{cccccccccc}

  \hline\hline
   Zone & $v_{\rm bulk}/c$ & $\sigma_v$~(km~s$^{-1}$) & $N_{\rm H}$~($10^{22}$~cm$^{-2}$)& $\log \xi$ & $C$-stat/d.o.f. & $\Delta$$C$-stat/$\Delta$d.o.f. & AIC & $\Delta$AIC & Relative $\mathcal{L}$ \\\hline
   $-$ & $-$& $-$&$-$& $-$ &  8014.33/9884 & $-$ & 8032.34 & $-$ \\
   1 & $0.076 \pm 0.001$ & $822^{+885}_{-326}$ & $8.7^{+5.0}_{-4.0}$ & $3.98^{+0.27}_{-0.25}$  &  7991.27/9880 & $-23.06/4$ & 8017.31 & $-15.03$ & $5.4\times10^{-4}$ \\
   2 & $0.101^{+0.002}_{-0.001}$ & $4227^{+1602}_{-1408}$ & $26.7^{+16.8}_{-13.7}$ & $3.91^{+0.26}_{-0.30}$  & 7933.58/9876 & $-57.69/4$ & 7967.64 & $-49.67$ & $1.6\times10^{-11}$\\
   3 & $0.143^{+0.005}_{-0.004}$ & $5795^{+2613}_{-3291}$ & $22.1^{+14.1}_{-6.1}$ & $4.12^{+0.59}_{-0.32}$  & 7917.07/9872 & $-16.51/4$ & 7959.16 & $-8.48$ & $0.014$ \\[1.5ex]
			
  \hline\hline
  \end{tabular}
\end{center}
\label{tab1}
\end{table*}

During the PV phase, XRISM observed IRAS~05189$-$2524 at $(\textrm{RA}, \textrm{Dec}) = (80.255831, -25.362589)$ from August 9 to 13, 2024 (UTC), with the net exposure time of 160~ksec (OBSID: 300020010). 
The two instruments are on board XRISM. 
One is the soft X-ray spectroscopy telescope, Resolve, composed of the X-ray microcalorimeter \citep{Ishisaki2022} and the X-ray Mirror Assembly (XMA; \citealt{Hayashi2024}).
The other is the soft X-ray imaging telescope, Xtend, which consists of the X-ray CCD camera \citep{Noda2025} and another XMA \citep{Tamura2024}. 
Resolve achieves the unprecedentedly high energy resolution of $\sim 5$~eV at 6~keV, while the Xtend achieves the wide field of view of $38.5 \times 38.5$ arcmin square over the 0.4--12~keV band, which is broader than that of Resolve when the gate valve is closed.
The data reduction was done by using heasoft-6.34 and CALDB as of November 2024. 

The Resolve X-ray microcalorimeter data were reduced using standard procedures. 
We extracted only the high-resolution primary (Hp) grade events and constructed the source spectrum by combining all pixels except for PIXEL 27, where large gain fluctuations are confirmed. 
The spectra were binned with the \texttt{grppha} command to ensure that each bin contained at least one event, and for visualization in the figures, they were further rebinned using the \texttt{setplot rebin} command. 
The Redistribution Matrix File (RMF) was generated using the \texttt{rslmkrmf} command with the whichrmf = X option, after removing the anomalous low-resolution secondary (Ls) grade events. 
After creating an exposure map with xaexpmap, we produced the Auxiliary Response File (ARF) using \texttt{xaarfgen}. 
Non X-ray Background (NXB) was extracted by the \texttt{rslnxbgen} program with the good time interval selection between the source and the NXB, and the model that reproduces the generated NXB spectrum is included in the Resolve spectral analyses. 
Since the gate valve was closed during this observation, we restricted the analysis to the 1.7--12.0~keV energy band. 
The Resolve spectrum was fitted in XSPEC using $C$-statistic.

The Xtend X-ray CCD data were derived with the full-window normal mode, and we reduced the data using the standard method for it. 
We extracted the source spectrum from a $120''$ circular region centered on the source and the background spectrum from an identical circular region located off-source on the same CCD chip. 
The RMF and ARF files were generated using the \texttt{xtdrmf} and \texttt{xaarfgen} commands, respectively. 
The source spectrum was binned with the grppha algorithm to ensure a minimum of 20 events per bin, and for display purposes in the figures, it was further rebinned with the \texttt{setplot rebin} command.
The fit to the Xtend spectra was performed in XSPEC using $\chi^2$-statistic.

\section{Data Analyses and Results}
\label{three}
\subsection{Resolve spectrum}
\label{3.1}

Figure~\ref{resolve_spec} shows the 5--10.5~keV Resolve spectrum of IRAS~05189$-$2524, and the most striking feature is the presence of multiple complex absorption lines in the $\sim$7--9~keV range. 
Additionally, a broad emission line-like feature can be seen around $\sim 7$~keV.
To characterize the spectral shape, we fitted the 1.7--12~keV Resolve spectrum using the model of XSPEC: \texttt{Abs$_{\rm gal}$*Abs$_{\rm int}$*$[$(multiple XSTAR$_{\rm UFO}$)*PL + Gauss$]$}.
Here, \texttt{PL} represents the thermal Comptonization continuum, modeled with \texttt{zpowerlaw}, and \texttt{Gauss} corresponds to the broad emission line near $\sim7$~keV, modeled with \texttt{zgaussian}. 
The neutral photoelectric absorption by Galactic and intrinsic matter is accounted for by multiplicative components \texttt{Abs$_{\rm Gal}$} and \texttt{Abs$_{\rm int}$}, modeled with \texttt{TBabs} and \texttt{zTBabs}, respectively.
To reproduce the complex absorption structures seen in $\sim7$--9~keV, we introduced multiple photoionized absorption components (\texttt{XSTAR$_{\rm UFO}$}) generated with the XSTAR code \citep{Kallman2004}. 
The spectral energy distribution (SED) used in the XSTAR calculation assumes a power-law shape, with the photon index tied to that of the \texttt{PL} component.

In the spectral fitting, the redshift ($z$) was fixed at $0.0426$ for all components except where explicitly stated.
The photon index ($\Gamma$) and normalization ($N_{\rm PL}$) of the \texttt{PL} component were left free, as were the central energy ($E$), width ($\sigma$), and normalization ($N_{\rm Gauss}$) of the \texttt{Gauss} component.
The column density ($N_{\rm H}$) of \texttt{Abs$_{\rm Gal}$} was fixed at $2 \times 10^{20}$~cm$^{-2}$, while that of \texttt{Abs$_{\rm int}$} was left free.
For each \texttt{XSTAR$_{\rm UFO}$} component, the bulk velocity ($v_{\rm bulk}$), velocity dispersion ($\sigma_{v}$), column density ($N_{\rm H}$), and ionization parameter ($\xi$) were all treated as free parameters.
We first performed the fit without any \texttt{XSTAR$_{\rm UFO}$} components, using the model \texttt{Abs$_{\rm gal}$*Abs$_{\rm int}$*(PL + Gauss)}.
This resulted in a $C$-statistic/d.o.f of 8014.33/9884, as summarized in Table~\ref{tab1}.

\begin{figure*}[t]
\epsscale{1.1}
\plotone{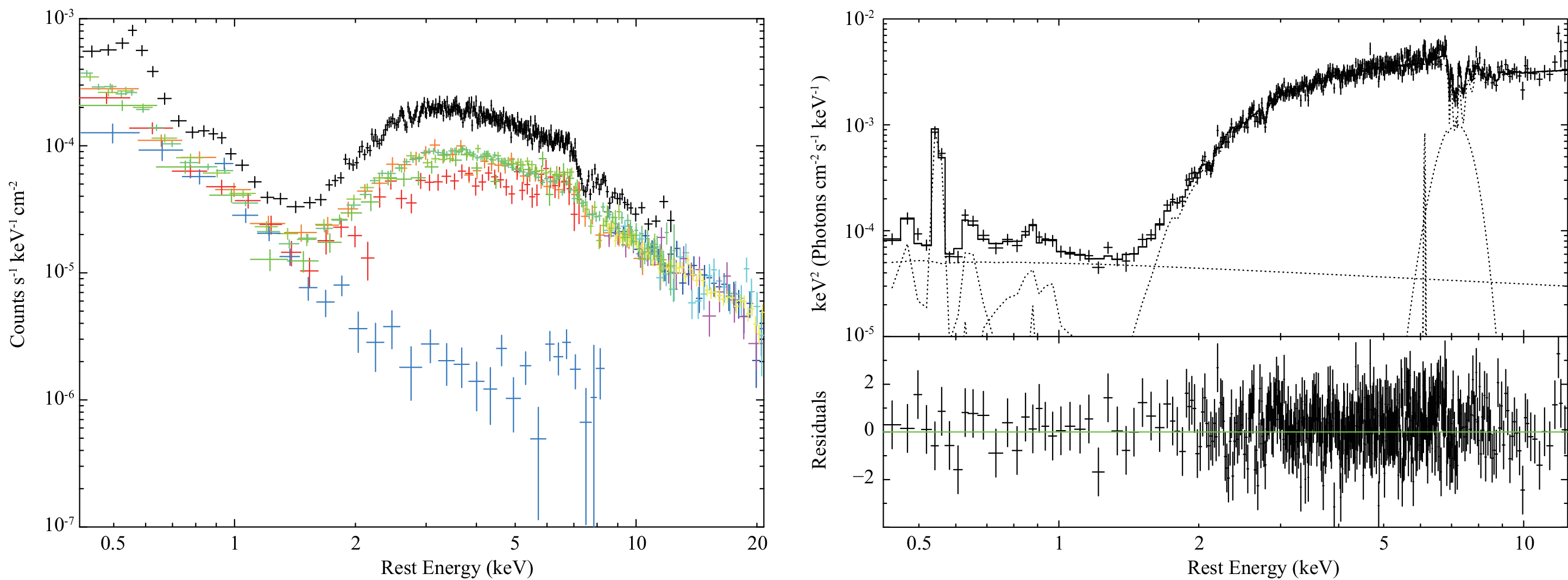}
\caption{(Left) The Xtend spectrum normalized with the effective area (black) shown with the spectra obtained by Chandra/ACIS (red and green), XMM-Newton/EPIC PN (orange, light green, and yellow-green), Suzaku/XIS (light blue), and NuSTAR (blue, cyan, magenta, and yellow) which were used in \cite{Yamada2021}. (Right) The deconvolved Xtend spectrum in a $\nu F_{\nu}$ form fitted with the model of \texttt{Abs$_{\rm Gal}$*$[$Abs$_{\rm int}$*$[$(three XSTAR$_{\rm UFO}$)*PL + Gauss + Fe-K$\alpha$$]$ + PL$_{\rm scat}$ + HostGal$]$}. Both are shown in the observer frame.}
\label{xtend_spec}
\end{figure*}

Subsequently, we added the \texttt{XSTAR$_{\rm UFO}$} components one by one, ultimately incorporating a total of three absorption zones.
Table~\ref{tab1} presents the best-fit parameters for each \texttt{XSTAR$_{\rm UFO}$} zone, along with the corresponding improvements in the $C$-statistic, the Akaike Information Criterion (AIC; \citealt{Akaike1974}), and the relative likelihood (Relative $\mathcal{L}$) associated with the inclusion of each component.
As a result, Zones~1 and 2 were found to be statistically significant, while Zone 3 was marginal based on the AIC criterion. 
We thus identify two significant UFO components, with bulk velocities of $v_{\rm bulk}/c \sim 0.076$ and 0.101.
Because \cite{Smith2019} reported the component with $v_{\rm bulk}/c \sim 0.1$, the component with $v_{\rm bulk}/c \sim 0.075$ represents a previously unreported feature, and was detected here for the first time thanks to the high-resolution capability of Resolve.
The velocity dispersion also varies among the components. 
Zone 1 with $v_{\rm bulk}/c \sim 0.076$ produces a relatively narrow absorption line with $\sigma_{v} \sim 800$~km~s$^{-1}$, whereas Zone 2 with $v_{\rm bulk}/c \sim 0.101$ yields a much broader absorption feature with $\sigma_{v} \sim 4200$~km~s$^{-1}$. 
This suggests a coexistence of both narrow and broad absorption structures around $\sim 7$--9~keV. 
It should be noted that, unlike in PDS~456 where each zone is dominated by a single Fe~XXV line \citep{XRISM2025}, both Fe~XXV and Fe~XXVI lines are observed in IRAS~05189$-$2524.
Although absorption-like residuals can be seen around $\sim 7.0$~keV and $\sim 9.1$~keV in the rest frame, the corresponding UFO components are not significantly detected. 
Additional observations with XRISM will be essential to investigate these features in more detail.
As shown in Table~\ref{tab1}, the $C$-statistic and AIC values decrease with the inclusion of each of the three UFO components, and hence, we decide to include all three UFO components in the final best-fit model. 
After accounting for the three-zone UFO components, the best-fit model is shown in Figure~\ref{resolve_spec}(a).
The photon index of the \texttt{PL} component became $\Gamma = 2.29^{+0.09}_{-0.05}$, indicating a steep power-law continuum. 
Regarding the \texttt{Gauss} component, the best-fit values for the energy, width, and normalization were found to be $E = 7.39^{+0.15}_{-0.17}$~keV, $\sigma = 537^{+168}_{-107}$~eV, and $N_{\rm Gauss} = (3.18^{+0.78}_{-0.80}) \times 10^{-5}$ photons~cm$^{-2}$~s$^{-1}$, respectively. 

Although the \texttt{Gauss} component could, in principle, produce a characteristic P~Cygni profile in combination with the UFO absorption features, the best-fit profile is found to be largely centered on the UFO absorption systems. To further examine a self-consistent P~Cygni scenario, we constructed an emission model for the UFOs using the XSTAR code, following the same approach as in \cite{XRISM2025}.
In this model, we assumed a velocity range uniformly distributed between 0.076$c$ and 0.143$c$. 
The column density was tied to the sum of those in Zones~1, 2, and 3, and the ionization parameter was tied to that of Zone 2, which is the most significant absorption component. 
We replaced the \texttt{Gauss} component with the XSTAR emission model, leaving the inclination angle and the normalization free, which reduces the number of free parameters by 1 from the \texttt{Gauss} model.
We fitted the 1.7--12~keV Resolve spectrum and obtained a successful fit, as shown in Figure~\ref{resolve_spec}(b). The resulting $C$-stat/d.o.f. and AIC values were 7923.05/9873 and 7963.14, respectively, which are marginally worse than those reported in Table~\ref{tab2}.
This may indicate the presence of components faster than the UFOs, which exhibit significant absorption lines.

Even when the \texttt{Gauss} component was replaced by the XSTAR emission model, all parameters of the UFO absorption lines remained consistent with those in Table~\ref{tab1} within the statistical uncertainties.
The inclination angle was poorly constrained, with a best-fit value of $48.1^{+20.8}_{-31.6}$ degrees. 
The normalizaion of the emission line became $0.5^{+1.1}_{-0.1} \times 10^{-3}$.  
The normalization of the model is defined as $K= (F_{\rm cov}/2)(L_{\rm ion}/10^{38}~\textrm{erg~s}^{-1})/(D/\textrm{kpc})^2 \sim 6 \times 10^{-5}~F_{\rm cov}$, where $F_{\rm cov}$ is the covering fraction of the UFOs, $L_{\rm ion}$ is the ionizing luminosity, and $D_{\rm }$ is the distance to the source \citep{XRISM2025}.
Here, we adopted $L_{\rm ion} = 4.1 \times 10^{44}$~erg~s$^{-1}$, calculated as the luminosity of the unabsorbed power-law component integrated over 1--1000~Ryd (see \S3.2).
Therefore, the derived normalization is 5 times higher than expected for $F_{\rm cov} = 1$, even considering the lower bound of the error range. 
This suggests that the actual covering fraction may be close to $F_{\rm cov} \sim 1$. 
The apparently unphysical result of $F_{\rm cov} > 1$ is likely due to the large uncertainty in the ionizing luminosity, contributions from UFO components that are not significantly detected, and/or reprocessing by the inner region of the accretion disk that is not fully accounted for in the model.

To quantitatively test the case in which neither the \texttt{Gauss} nor the XSTAR emission model is included, we excluded both components and fitted the spectrum with \texttt{Abs$_{\rm gal}$*Abs$_{\rm int}$*(multiple XSTAR$_{\rm UFO}$)*PL}. 
Figure~\ref{resolve_spec}(c) shows the fitting results. 
The resulting $C$-stat/d.o.f. and AIC values were 7977.93/9875 and 8014.00, respectively, indicating that the fit was significantly worse than fits including the \texttt{Gauss} or XSTAR emission model. 
Positive, broad emission line-like residuals can be seen from $\sim 6.5$ to $\sim 8$~keV in Figure\ref{resolve_spec}(c), suggesting that the broad emission line is indeed present as part of a P~Cygni profile produced by the UFOs. 
In this fit, all parameters of the UFO absorption lines remained consistent with those in Table~\ref{tab1} within the statistical uncertainties, suggesting that the absorption lines are not strongly affected by the presence of the broad emission line.

Finally, we investigated the presence of a neutral Fe-K$\alpha$ emission line around 6.4~keV.
We added \texttt{Fe-K$\alpha$}, which consists of two \texttt{zgauss} components for the Fe-K$\alpha_1$ and Fe-K$\alpha_2$ lines, to the best-fit model, with their line energies fixed at 6.404~keV and 6.391~keV, respectively, as shown in Figure~\ref{resolve_spec} (a). 
Therefore, the finally fitted model is \texttt{Abs$_{\rm Gal}$*Abs$_{\rm int}$*$[$(multiple XSTAR$_{\rm UFO}$)*PL + Gauss + Fe-K$\alpha$$]$}. 
The normalization of the Fe-K$\alpha_1$ component was fixed at twice that of the Fe-K$\alpha_2$ component.
The line widths ($\sigma$) and normalizations ($N_{\rm Gauss}$) were tied between the two components and left free in the fit.
To account for the possibility that the Fe-K$\alpha$ emitter is slightly ionized, we allowed the redshift ($z$) to vary freely.
As a result, the \texttt{Fe-K$\alpha$} component was found to be marginally required, with $\Delta$$C$-statistic/$\Delta$d.o.f.$ = -9.53/3$ and $\Delta$AIC$ = -3.51$.
The best-fit Gaussian width was $\sigma = 16.1^{+16.7}_{-14.5}$~eV, corresponding to a velocity width of 180--3600~km~s$^{-1}$ (FWHM), and the equivalent width was $15^{+14.2}_{-11.4}$~eV.
Although the velocity width is not well constrained, the equivalent width was confirmed to be smaller than that of typical Seyfert galaxies.
The best-fit parameters to the Resolve spectrum, except for the UFO components already shown in Table~\ref{tab1}, are summarized in Table~\ref{tab2}.

\subsection{Xtend spectrum}
\label{3.2}

Figure~\ref{xtend_spec} (left) shows the the 0.4--12.0~keV Xtend spectrum, shown together with the X-ray spectra by previous observations by Chandra, XMM-Newton, Suzaku, and NuSTAR which were reported by \cite{Yamada2021}. 
This indicates that XRISM successfully captured IRAS~05189-2514 during one of its historically bright phases.
The absorption feature caused by the UFOs, which were investigated in detail through the spectral fit with Resolve, can also be clearly seen around 7.5~keV in the Xtend spectrum. 
In order to investigate the broadband spectral properties, we performed the spectral fit to the 0.4--12.0~keV Xtend spectrum.

\renewcommand{\arraystretch}{1}
\begin{table}
 \caption{Parameters of the fits to the Resolve 1.7--12~keV and Xtend 0.4--12~keV spectra, except for those of multiple \texttt{XSTAR$_{\rm UFO}$} components.}
 \label{tab2}
 \begin{center}
 \begin{tabular}{cccc}

  \hline\hline
   Model & Parameter & Resolve & Xtend \\\hline
   \texttt{Abs$_{\rm Gal}$} & $N_{\rm H}^{a}$~($10^{20}$~cm$^{-2}$) & 2~(fix) &2~(fix) \\[1.5ex]
   \texttt{Abs$_{\rm int}$} & $N_{\rm H}^{a}$~($10^{22}$~cm$^{-2}$) & $7.3^{+0.9}_{-1.0}$ & $7.0 \pm 0.2$ \\[1.5ex]
 \texttt{XSTAR$_{\rm UFO} \times 3$} & \multicolumn{3}{c}{= Table~\ref{tab1}} \\[1.5ex]	
    \texttt{PL} & $\Gamma$ & $2.29^{+0.09}_{-0.05}$ &  $2.22 \pm 0.05$ \\
    			& $N_{\rm PL}^{b}$ & $0.80 \pm 0.13$ & $ 0.71 \pm 0.06$ \\[1.5ex]				
    \texttt{Gauss} & $E$~(keV) & $7.39^{+0.15}_{-0.17}$ & 7.39~(fix) \\
    			 & $\sigma$~(eV) & $537^{+168}_{-107}$ & 537~(fix) \\
			 & $N_{\rm Gauss}^{c}$ &  $3.18^{+0.78}_{-0.80} $ & 3.18~(fix) \\
			 & EW~(eV) & $813^{+289}_{-321}$ & 813~(fix) \\[1.5ex] 
    \texttt{Fe-K$\alpha$} & $\sigma$~(eV) & $16.1^{+16.7}_{-14.5}$ & 16.1~(fix) \\
    			& EW~(eV) & $15.0^{+14.2}_{-11.4}$ & 15.0~(fix) \\[1.5ex]
     \texttt{PL$_{\rm scat}$} & $N_{\rm PL}^{d}$ & $-$ & $5.69^{+0.72}_{-0.73}$  \\[1.5ex]		
    \texttt{HostGal}  & $kT_1$~(keV) & $-$ & $0.86 \pm 0.11$ \\
			    & $N_{\rm Host1}^{e}$ & $-$ & $0.10 \pm 0.04$ \\
			    & $kT_2$~(keV) & $-$ & $0.12 \pm 0.02$ \\
			     & $N_{\rm Host2}^{e}$ & $-$ & $3.06^{+2.08}_{-1.10}$ \\\hline
	\multicolumn{2}{c}{$C$-stat/d.o.f. or $\chi^2$/d.o.f.}  & 7917.07/9872 &1351.79/1305\\		     
  \hline\hline
  \end{tabular}
\end{center}
\tablenotetext{a}{Equivalent hydrogen column density.}
\tablenotetext{b}{Normalization of \texttt{zpowerlaw} in $10^{-2}$~photons~keV$^{-1}$~cm$^{-2}$~s$^{-1}$ at 1~keV.}
\tablenotetext{c}{Normalization of \texttt{zgaussian}  in $10^{-5}$ photons~cm$^{-2}$~s$^{-1}$.}
\tablenotetext{d}{Normalization of \texttt{zpowerlaw} in $10^{-5}$~photons~keV$^{-1}$~cm$^{-2}$~s$^{-1}$ at 1~keV.}
\tablenotetext{e}{Normalization of \texttt{apec} in $\frac{10^{-18}}{4\pi [D_A (1+z)]} \int n_{\rm e} n_{\rm H} dV$, where $D_A$ is the angular diameter distance to the source, $dV$ is the volume element, and $n_{\rm e}$ and $n_{\rm H}$ are electron and hydrogen densities, respectively. }
\label{tab2}
\end{table}

As expected for a Compton-thin type-2 AGN, the soft X-ray continuum of IRAS~05189$-$2524 is heavily absorbed due to obscuration by neutral material.
As reported by \cite{Yamada2021}, the soft X-ray band appears to be dominated by thin thermal plasma emission originating from the host galaxy and the AGN scattered continuum.
In the 6--8~keV band of the Xtend spectrum, complex absorption features, as described in Section~\ref{3.1}, are also evident.
Therefore, we assumed the model of 
\texttt{Abs$_{\rm Gal}$*$[$Abs$_{\rm int}$*$[$(three XSTAR$_{\rm UFO}$)*PL + Gauss + Fe-K$\alpha$$]$ + PL$_{\rm scat}$ + HostGal$]$}.
\texttt{Abs$_{\rm Gal}$}, \texttt{Abs$_{\rm int}$}, and \texttt{PL} are the same as those in the fit to the Resolve spectrum (\S3.1).
\texttt{PL}$_{\rm scat}$ represents the AGN scattered continuum, modeled with \texttt{zpowerlw}, where the photon index was tied to that of \texttt{PL}, and the normalization was left free.
\texttt{HostGal} represents the thin thermal emission from the host galaxy, modeled with two \texttt{apec} components. 
The plasma temperatures and normalizations of the two \texttt{apec} components were left free and untied, while their metal abundances were fixed at the Solar value.
The parameters of the three \texttt{XSTAR$_{\rm UFO}$} components, as well as those of the \texttt{Gauss} and \texttt{Fe-K$\alpha$} lines, were fixed to the best-fit values obtained from the fit to the Resolve spectrum in Table~\ref{tab2}.
Under these assumptions, we performed spectral fitting of the Xtend 0.4--12.0~keV data.

Figure~\ref{xtend_spec} and Table~\ref{tab2} show the best-fit parameters. 
We successfully reproduced the Xtend broadband spectrum. 
The obtained column density ($N_{\rm H}$) of the neutral absorption and the photon index ($\Gamma$) of \texttt{PL} were consistent with those derived in the Resolve 1.7--12.0~keV within errors. 
The observed 2--10~keV flux without correcting absorption was $F_{2-10} = 6.7 \times 10^{-12}$~erg~cm$^{-2}$~s$^{-1}$ which is $\sim50$\% higher than that derived by XMM-Newton and NuSTAR in 2013 (\citealt{Xu2017}; \citealt{Smith2019}).
 After correcting for absorption, the ionizing luminosity ($L_{\rm ion}$), calculated as the luminosity of the unabsorbed power-law component integrated over 1--1000~Ryd, is $L_{\rm ion} = 4.1 \times 10^{44}$~erg~s$^{-1}$.
Note that the value of $L_{\rm ion}$ presented here corresponds to a lower limit, assuming a simple power-law spectral energy distribution.
For example, when including a disk blackbody component corresponding to an accretion disk radiating at an Eddington ratio of unity in addition to the power-law, the ionizing luminosity increases to $L_{\rm ion} \sim 2.7 \times 10^{45}$~erg~s$^{-1}$.
The inclusion of a soft X-ray excess component may further enhance this value.

\begin{figure}[t]
\epsscale{1.2}
\plotone{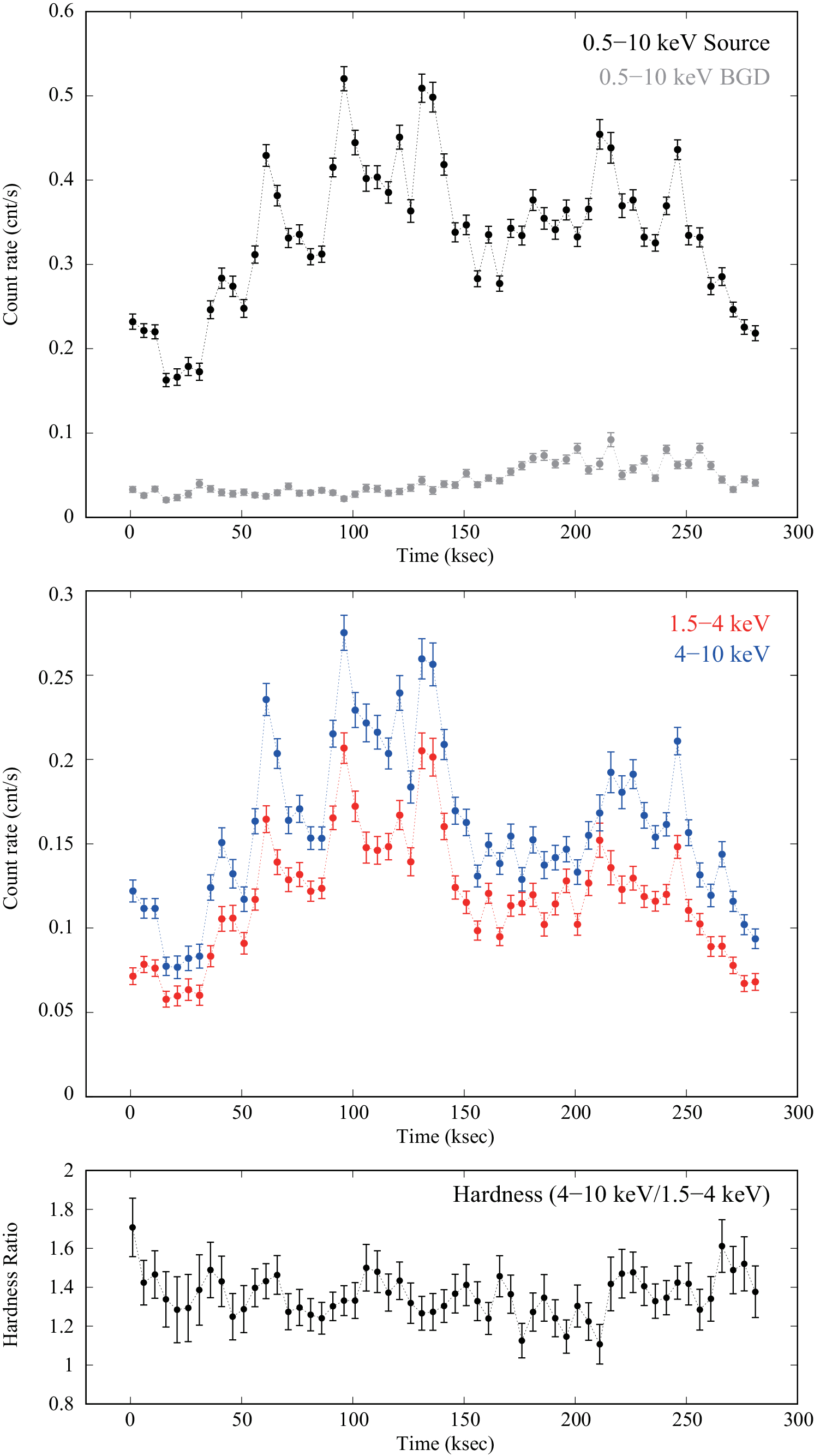}
\caption{(Top) The 0.5--10~keV source (black) and background (grey) light curves derived by Xtend. (Middle) The 1.5--4~keV (red) and 4--10~keV (blue) light curves. (Bottom) The hardness ratio obtained by dividing the 4--10~keV count rate by the 1.5--4~keV count rate. The observation start time is 2024 August 9 17:34:01 (UT). }
\label{xtend_lc}
\end{figure}

\subsection{Light curves and the UFO variability}
\label{3.3}

\begin{figure}[t]
\epsscale{1.1}
\plotone{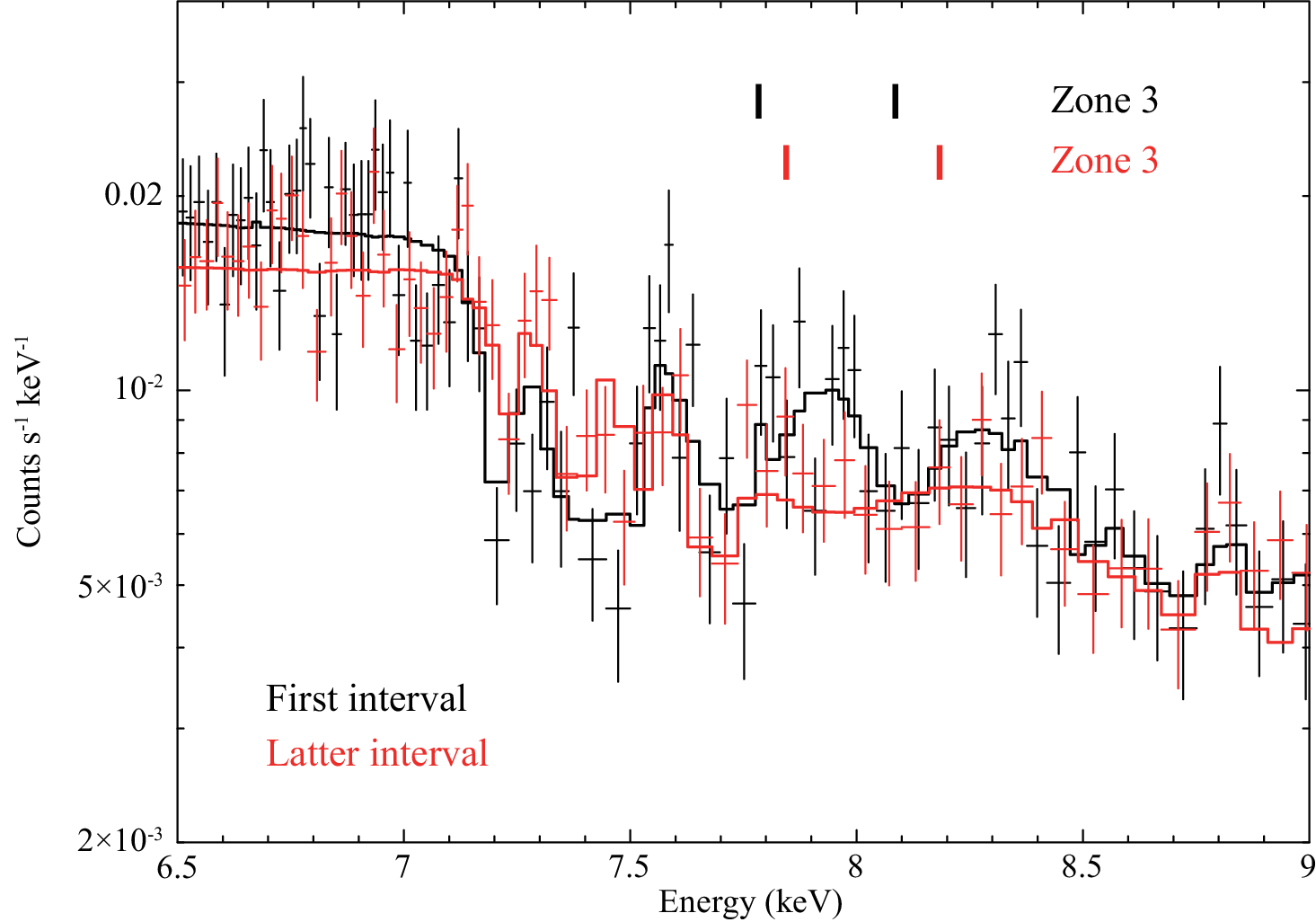}
\caption{Resolve spectra in the rest frame, divided into the first 140~ksec (black) and the latter 144~ksec (red). The black and red lines show the best-fit models for the first and latter intervals, respectively, obtained from a simultaneous fit using the model \texttt{Abs$_{\rm Gal}$*Abs$_{\rm int}$*[$($three XSTAR$_{\rm UFO}$)*PL + Gauss + Fe-K$\alpha$]}, in which the UFO parameters were untied and allowed to vary freely.
Vertical solid lines indicate the approximate positions of the Fe XXV/XXVI K$\alpha$ absorption lines in Zone~3 at the first interval (black) and the latter interval (red). }
\label{UFO_variability}
\end{figure}

Figure~\ref{xtend_lc} (top) presents the 0.5--10~keV light curve of IRAS~05189$-$2524 obtained with Xtend. 
To evaluate the significance of the observed variability, a background light curve extracted from the same energy band and a region of comparable area is also shown for comparison.
Although the background count rate exhibits some increase in the later portion of the observation, the variability amplitude of IRAS~05189$-$2524 is approximately 5--10 times larger, indicating that the variability of the AGN is significant.
Around 60~ks after the start of the observation, a series of flares exhibiting sudden flux variations of roughly 20--30\% becomes apparent. 
Each flare has a duration of roughly 10~ksec, and flares appeared intermittently with separations of about 40~ksec.

Next, to examine whether the observed variability is caused by intrinsic changes in the X-ray emission of the AGN or by variations in the column density of the neutral absorber, we extracted light curves in the 1.5--4~keV band where neutral absorption is dominant and the 4--10~keV band where the hard X-ray emission from the AGN dominates, as shown in Figure~\ref{xtend_lc} (middle).
We found that both bands exhibit similar variability patterns, with prominent flares clearly detected in each.
Figure~\ref{xtend_lc} (bottom) presents the hardness ratio, defined as the 4--10~keV count rate divided by the 1.5--4~keV count rate.
As a result, no significant variation in the hardness ratio was observed throughout the observation.
These results indicate that the drastic variability is primarily due to intrinsic changes in the X-ray emission from the AGN corona.

As shown in Figure~\ref{xtend_lc}, pronounced flare-like variability is observed in the first interval, whereas the latter interval appears relatively quiescent.
To investigate the variability of the UFO components between these intervals, we divided the Resolve data into the first 140~ksec and the latter 144~ksec, and extracted the 1.7--12~keV spectra.
We simultaneously fitted both spectra with the model \texttt{Abs$_{\rm Gal}$*Abs$_{\rm int}$*$[$(three XSTAR$_{\rm UFO}$)*PL + Gauss + Fe-K$\alpha$}$]$, which is identical to the best-fit model described in \S3.1.
First, the parameters $v_{\rm bulk}$, $\sigma_v$, $N_{\rm H}$, and $\log \xi$ of the three \texttt{XSTAR$_{\rm UFO}$} components were tied, while the power-law normalization was left untied, and all other model parameters were tied. The resulting $C$-statistic was 10238.40 with d.o.f. $ = 12764$. Next, we allowed all \texttt{XSTAR$_{\rm UFO}$} parameters in all zones to vary independently between the two spectra, yielding a $C$-statistic of 10206.19 with d.o.f. $ = 12752$. 
This improvement of $\Delta C = 32.21$ for 12 degrees of freedom suggests possible variability in the UFO parameters.
Figure~\ref{UFO_variability} shows the best-fit model when all \texttt{XSTAR$_{\rm UFO}$} parameters were untied.
To identify which zone exhibits variability, we compared the 90\% confidence intervals of the parameters for each zone.
As a result, no statistically significant changes were found for the UFO parameters in Zones~1 and 2.
Zone~3 exhibited possible indications of variability, with  $v_{\rm bulk}$, $\log \xi$, and $\sigma_v$ not overlapping within the 90\% confidence ranges. 
Specifically,  $v_{\rm bulk}$ increased from $(0.103 \pm 0.005)c$ to $(0.114^{+0.002}_{-0.001})c$, $\log \xi$ decreased from $4.17^{+0.70}_{-0.34}$ to $3.48^{+0.23}_{-0.19}$, and $\sigma_v$ increased from $4661^{+3177}_{-1295}$~km~s$^{-1}$ to $>8464$~km~s$^{-1}$.
However, given that Zone~3 is only marginally detected in the time-averaged spectral fit (see \S3.1), this result should be interpreted with caution.
Further observations with higher signal-to-noise data will be required to confirm whether the UFO components truly exhibit variability on the timescale of $\sim 140$~ksec.

\section{Discussion}
\label{four}

\subsection{Summary of the results}

Here, we summarize the main results obtained by XRISM observation during the PV phase for IRAS~05189$-$2524:
\begin{itemize}
\item We discovered, for the first time, at least three zones of UFOs with multi-velocities and velocity dispersions from the nearly type-2 AGN in the merging galaxy with intense starburst activity, exhibiting blueshifted velocities of approximately $0.076c$, $0.101c$, and $0.143c$.
\item A broad emission line was detected around $\sim 7$~keV, with a velocity width consistent with the bulk velocities of the UFO components and an equivalent width of $\sim 813$~eV. 
\item A neutral Fe-K$\alpha$ emission line at $\sim 6.4$~keV was weak, with an equivalent width of $\sim 15$~eV, and its line width was loosely constrained to $\sim 180$--3600~km~s$^{-1}$ (FWHM). 
\item The primary X-ray continuum from the AGN is relatively steep, with a photon index of $\Gamma \sim 2.3$, and is absorbed by neutral material with a column density of $N_{\rm H} \sim 7 \times 10^{22}$~cm$^{-2}$.
\item Flare-like variations in the intrinsic hard X-ray emission from the AGN were observed, each lasting approximately $\sim10$~ksec and occurring intermittently at intervals of $\sim40$~ksec.
\item We detected potential variability in the UFO component with the highest bulk velocity between the first and latter intervals, on a timescale of $\sim140$~ksec.
\end{itemize}

\subsection{Multi-velocity UFOs from the AGN in the starburst merging galaxy}
\label{sec4.2}

One of the most important results of this study is the discovery that powerful UFO components are already present in a galaxy merger system, which represents a precursor stage before the emergence of a luminous quasar.
In the widely accepted scenario for quasar evolution proposed by \cite{Sanders1988} and \cite{Hopkins2006}, gas-rich disk galaxies merge, driving large amounts of gas toward the galactic center via gravitational torques, thereby triggering intense starburst activity.
Subsequently, accretion onto the central SMBH becomes more active, leading to strong AGN feedback that rapidly quenches the ongoing starburst.
According to simulations by \cite{Hopkins2006} and \cite{Hopkins2008}, the duration in which both strong AGN feedback and intense starburst activity coexist is expected to be short, on the order of tens Myr.
IRAS~05189$-$2524 appears to be in this transitional phase.
Therefore, the detection of powerful UFOs, likely manifestations of AGN feedback, during this rare evolutionary stage represents an observationally significant and timely discovery.

Another significant finding of this study is that the UFO components with multi-velocity and velocity dispersions were discovered in a nearly type-2 AGN. 
In the same PV phase of XRISM, the observation of the type-1 quasar PDS~456, which is well known to host UFOs, also revealed that the absorption features due to UFOs are not characterized by a single broad component, but rather by five distinct narrow absorption lines with different blueshifted velocities \citep{XRISM2025}.
The finding that IRAS~05189$-$2524, a nearly type-2 AGN, exhibits a similar set of multiple blueshifted absorption lines suggests that UFOs are not confined solely to small inclination angles, but rather exhibit a clumpy structure with density inhomogeneities that extend from the polar regions to the equatorial plane.
This inferred distribution may differ from those predicted by radiation magnetohydrodynamic simulations (e.g., \citealt{Takeuchi2013}), which do not produce clumpy structures in equatorial regions.

\subsection{Eddington ratio of the AGN in IRAS~05189$-$2524}

Because the AGN in IRAS~05189$-$2524 is classified as nearly type-2, direct optical--UV emission from the accretion disk and broad optical emission lines from the broad-line region are obscured, making the SMBH mass estimation highly uncertain.
Consequently, reliably determining the Eddington ratio has remained challenging; for instance, \cite{Teng2015} reported an Eddington ratio of 1.2, whereas \cite{Xu2017} estimated it to be as low as 0.12.
However, XRISM observations reveal a steep primary Comptonization continuum with a photon index of $\Gamma \sim 2.3$.
Furthermore, the X-ray emission exhibits significant variability, including multiple flares on timescales of $\sim10$~ksec.
If the 10~ksec flare duration is interpreted as the light-crossing time of the X-ray corona, the upper limit on the coronal size is estimated to be $\sim2$--$40~R_{\rm S}$ where $R_{\rm S}$ is the Schwarzschild radius, assuming a SMBH mass range from $M_{\rm BH} = 2.5 \times 10^7$ to $4.2 \times 10^8~M_{\odot}$.
This suggests that the X-ray corona is indeed compact. 
These properties are indicative of an accretion rate approaching the Eddington limit (e.g., \citealt{Kubota2018}).

The detected narrow Fe-K$\alpha$ emission line has an equivalent width of $\sim15$~eV, which is significantly smaller than the typical value observed in sub-Eddington Seyfert galaxies ($\sim100$~eV). According to studies of the X-ray Baldwin effect, the Fe-K$\alpha$ equivalent width decreases with increasing Eddington ratio, and such a low equivalent width of $\sim 15$~eV likely corresponds to an Eddington ratio of $\gtrsim1$ (e.g., \citealt{Ricci2013}).
Therefore, the Eddington ratio of IRAS~05189$-$2524 is likely close to unity, suggesting that the presence of UFO components with multiple velocities and velocity dispersions may be linked to mass accretion occurring near the Eddington limit onto the SMBH.
According to \cite{Ricci2017}, the covering fraction of the dusty torus decreases as the Eddington ratio approaches unity. This is consistent with the small equivalent width of the narrow Fe-K$\alpha$ line observed. If the neutral absorber is associated with the torus, then the observed $N_{\rm H}$ would require that this system be viewed from a nearly edge-on inclination.

\subsection{New constraints on the structure of UFOs with high-resolution X-ray spectroscopy}

\begin{figure*}[t]
\epsscale{0.9}
\plotone{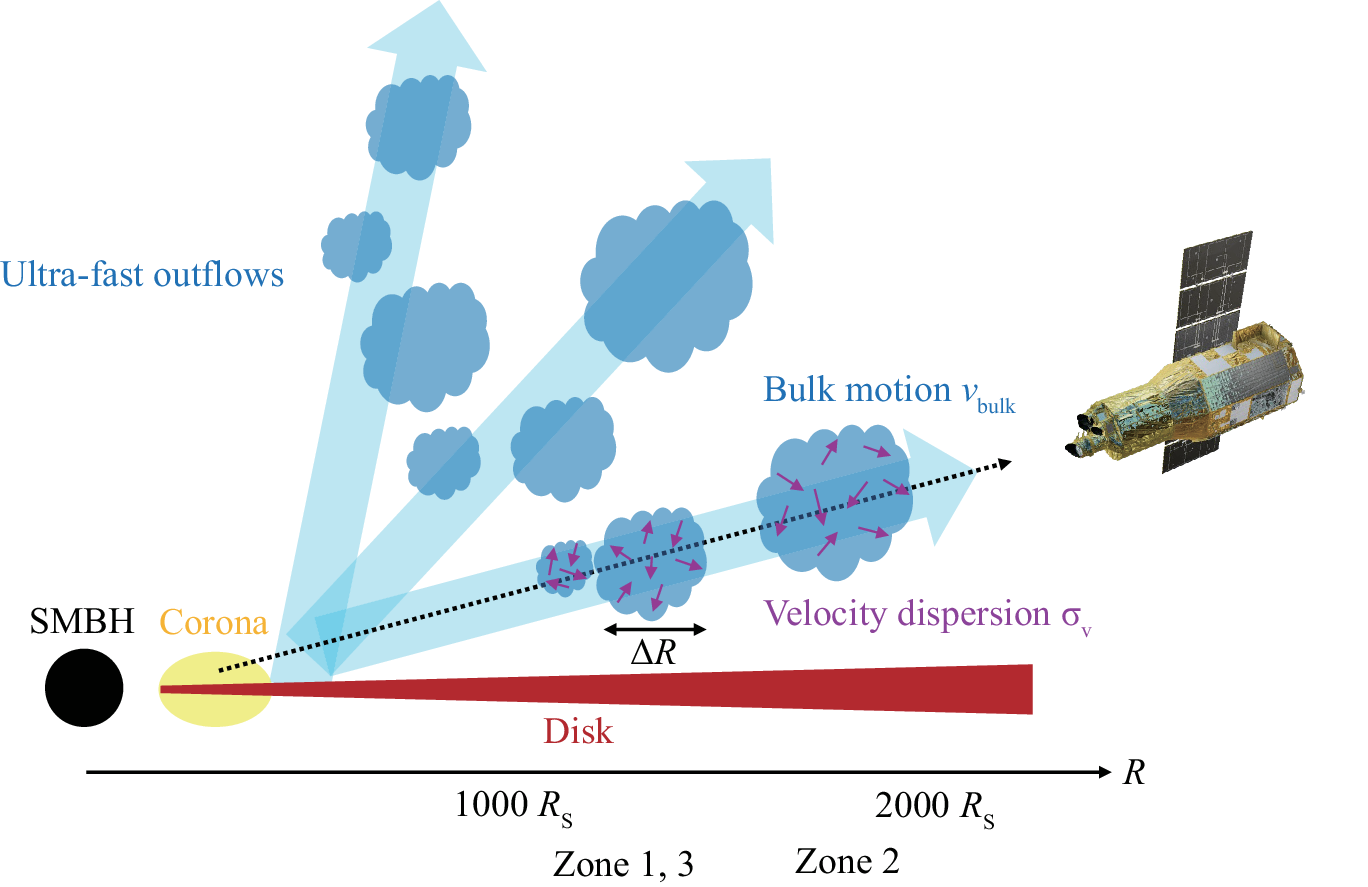}
\caption{Schematic picture of the UFO components with multi velocities and velocity dispersions in IRAS~05189$-$2524.}
\label{picture}
\end{figure*}

We identified multi-velocity UFO components (or clumps) along the line of sight and successfully constrained the physical properties of individual clumps, including their column density ($N_{\rm H}$), ionization parameter ($\xi$), ionizing luminosity ($L_{\rm ion}$), bulk velocity ($v_{\rm bulk}$), and velocity dispersion ($\sigma_v$).
Denoting the clump distance from the central source as $R$ and the clump number density as $n$, the ionization parameter is defined as $\xi = L_{\rm ion}/(n R^2)$. 
If the clump size is $\Delta R$, then $n = N_{\rm H}/\Delta R$, and substitution into the definition of $\xi$ yields
\begin{equation}
\frac{\Delta R}{R^2} = \frac{\xi N_{\rm H}}{L_{\rm ion}}~~~.
\label{eq1}
\end{equation}
We also found the potential variability of the UFO parameters in Zone~3 on a timescale of $\Delta t_{\rm UFO} \sim 140$~ksec (Section~\ref{3.3}). 
Interpreting the variability timescale as the clump transit time of the UFO clump yields an approximate clump size of
\begin{eqnarray}
\Delta R &\mathrel{\sim}& v_{\rm bulk} \Delta t_{\rm UFO}  \nonumber \\
 &\mathrel{\sim}& 80 \times \left(\frac{v_{\rm bulk}}{0.1c}\right)\left(\frac{\Delta t_{\rm UFO}}{200~\textrm{ksec}}\right)~R_{\rm S}~~~, 
 \label{eq2}
\end{eqnarray}
where $R_{\rm S}$ is the Schwarzschild radius corresponding to $M_{\rm BH} = 2.5 \times 10^7~M_{\odot}$.
When we substitte $v_{\rm bulk} \sim 0.14c$ and $\Delta t_{\rm UFO} \sim 140$~ksec, we obtain $\Delta R \sim 80~R_{\rm S}$ for Zone~3. 
Combining equations (\ref{eq1}) and (\ref{eq2}), the clump distance can be constrained as
\begin{eqnarray}
R &\mathrel{\sim}& \sqrt{\frac{L_{\rm ion}}{\xi N_{\rm H}} v_{\rm bulk} \Delta t_{\rm UFO}}  \nonumber\\
&\mathrel{\sim}& 1000 \times \left( \frac{L_{\rm ion}}{10^{44}~\textrm{erg~s}^{-1}}\right)^{\frac{1}{2}} \left( \frac{\xi}{10^{4}} \right)^{-\frac{1}{2}}  \nonumber \\
&& \times\left( \frac{N_{\rm H}}{10^{23}~\textrm{cm}^{-2}} \right)^{-\frac{1}{2}}  
\left(\frac{v_{\rm bulk}}{0.1c}\right)^{\frac{1}{2}} \left(\frac{\Delta t_{\rm UFO}}{200~\textrm{ksec}}\right)^{\frac{1}{2}}~R_{\rm S}~~~.
\label{eq3}
\end{eqnarray} 
Using the observed parameter values for Zone~3 and $L_{\rm ion} = 4.1 \times 10^{44}$erg~s$^{-1}$, we obtain $R \sim 1200~R_{\rm S}$. 
These results suggest that the clump of Zone~3, with a characteristic size of approximately a hundred $R_{\rm S}$ is located at a distance of about a thousand $R_{\rm S}$, i.e., in the vicinity of the central SMBH, as illustrated in Figure~\ref{picture}.

What is the physical origin of the UFO clumps exhibiting the observed velocity dispersions? 
In principle, there are two possible mechanisms. 
One possibility is that turbulent or velocity-shear structures with a characteristic scale of $\Delta R$ develop through radiative-hydrodynamic instabilities (e.g., \citealt{Takeuchi2013}; \citealt{Kobayashi2018}) as the outflow propagates to a distance $R$.
In this case, the clumps are expected to grow on the timescale of the time required for the velocity dispersion to traverse $\Delta R$, like the concept of the eddy turnover time, defined as 
\begin{equation}
t_{\rm traverse} = \frac{\Delta R}{\sigma_v}~~~.
\label{eq4}
\end{equation}
When the time required for the UFO component to travel from the center to a distance $R$ at a bulk velocity $v_{\rm bulk}$ is defined as
\begin{equation}
t_{\rm travel} = \frac{R}{v_{\rm bulk}} ~~~, 
\label{eq5}
\end{equation}
$t_{\rm traverse} \lesssim t_{\rm travel}$ is expected, because in the case of turbulence, the clump development timescale $t_{\rm traverse}$ must be shorter than the travel time $t_{\rm travel}$, whereas for velocity shear, the local velocity gradient $1/t_{\rm traverse}$ is expected to be comparable to the global velocity gradient $1/t_{\rm travel}$.
Another possibility is that turbulence or velocity shears are seeded at the time of the outflow launching from the accretion disk, and we are observing the residuals of these inhomogeneities after they decay during expansion. 
In this case, in order for the clumps not to be destroyed before reaching a distance $R$, the opposite condition to the previous case is required, namely $t_{\rm traverse} \gtrsim t_{\rm travel}$.
From equations (\ref{eq1}), (\ref{eq4}), and (\ref{eq5}) we obtain
\begin{eqnarray}
\label{eq6}
\frac{t_{\rm traverse}}{t_{\rm travel}} &=& \frac{v_{\rm bulk}}{\sigma_v} \frac{\xi N_{\rm H}}{L_{\rm ion}} R \\
\label{eq7}
&=& \frac{v_{\rm bulk}}{\sigma_v} \sqrt{\frac{\xi N_{\rm H}}{L_{\rm ion}} \Delta R}~~~.
\end{eqnarray}
By combining equation (\ref{eq2}) with the above relations, we can express the ratio $t_{\rm traverse}/t_{\rm travel}$ in terms of observable quantities as
\begin{eqnarray}
\frac{t_{\rm traverse}}{t_{\rm travel}} &\mathrel{\sim}& \frac{v_{\rm bulk}}{\sigma_v} \sqrt{\frac{\xi N_{\rm H}}{L_{\rm ion}}  v_{\rm bulk} \Delta t_{\rm UFO}} \nonumber\\
&\mathrel{\sim}& 1 \times  \left( \frac{v_{\rm bulk}}{0.1c} \right)^{\frac{3}{2}}  \left( \frac{\sigma_v}{2500~\textrm{km~s}^{-1}} \right)^{-1} \left( \frac{\xi}{10^{4}} \right)^{\frac{1}{2}} ~~~  \nonumber\\
&& \times \left( \frac{N_{\rm H}}{10^{23}~\textrm{cm}^{-2}} \right)^{\frac{1}{2}}  \left( \frac{L_{\rm ion}}{10^{44}~\textrm{erg~s}^{-1}}\right)^{-\frac{1}{2}} \left(\frac{\Delta t_{\rm UFO}}{200~\textrm{ksec}}\right)^{\frac{1}{2}}~~~.
\label{eq8}
\end{eqnarray}
When we substitute the variability timescale and observed parameter values for Zone~3 into equation (\ref{eq8}), we find $t_{\rm traverse}/t_{\rm travel} \sim 0.5$, which is of order unity, making it difficult to distinguish between the two scenarios. 
Nevertheless, future precise X-ray spectroscopy, by enabling measurements of both the physical parameters and temporal variability of individual UFO clumps, will provide a powerful means to probe the origin and acceleration mechanism of clumpy UFOs.

Because variability was not confirmed in Zones~1 and 2, equations (\ref{eq2}) and (\ref{eq3}) cannot be directly applied to estimate the clump size $\Delta R$ and distance $R$ for these zones. Instead, by assuming that $t_{\rm traverse}/t_{\rm travel} = 0.5$, as inferred for Zone 3, we can constrain $R$ and $\Delta R$ directly from the observables using equations (\ref{eq6}) and (\ref{eq7}), respectively.
Using the parameters of Zones~1, 2, and 3 (Table~\ref{tab1}) and adopting $L_{\rm ion} = 4.1 \times 10^{44}$erg~s$^{-1}$, we obtain distances of $R_1 \sim 1200~R_{\rm S}$, $R_2 \sim 1800~R_{\rm S}$, and $R_3 \sim 1300~R_{\rm S}$, respectively, for $M_{\rm BH} = 2.5 \times 10^7M_{\odot}$. These results indicate that the UFO components are distributed within $\sim 1800~R_{\rm S}$. 
The clump sizes are also estimated to be $\Delta R_1 \sim 20~R_{\rm S}$, $\Delta R_2 \sim 120~R_{\rm S}$, and $\Delta R_3 \sim 90~R_{\rm S}$.
Note that the constraints on both $R$ and $\Delta R$ are based on the assumption that the UFO varies on a timescale of $\Delta t_{\rm UFO} \sim 140$~ksec. The variability may occur on shorter timescales; therefore, the derived values should be regarded as upper limits.
Although Zones~1, 2, and 3 are located at similar distances of $\sim 1200$--$1800~R_{\rm S}$ from the SMBH, their velocity dispersions and region sizes differ by a factor of $\sim 6$. Such a discrepancy has not been reported in observations or simulations of other sources to date, making this an intriguing and noteworthy result.
This suggests that, rather than supporting a picture in which broader absorption lines are produced closer to the SMBH and narrower ones farther away, like the idea of escape velocity, a more natural interpretation is that regions with a wide range of velocity dispersions coexist at similar radii, providing a new observational constraint on the acceleration mechanisms of UFOs.

A schematic picture of the UFO components with multiple velocities and velocity dispersions, which is predicted from these results, is shown in Figure~\ref{picture}. 
Clumps with a range of sizes differing by a factor of $\sim 6$ and exhibiting velocity dispersion are thought to form regardless of the distance $R$ from the SMBH. 
For example, clumps such as Zone 2 and Zone 3, which exhibit a bulk velocity difference of $\sim 0.04c$ despite being separated by only $\sim 500~R_{\rm S}$, could easily generate internal shocks, suggesting that strong temporal variability in the UFO structure may occur on timescales of a month. 
The expected radius at which internal shocks occur is estimated to be $R_{\rm IS} \sim 3000~R_{\rm S}$. 
Such internal shocks can lead to various forms of dissipation (e.g., production of UFO neutrinos; \citealt{Wang2016}; \citealt{Liu2018}) and may also affect the global energetics of the outflow, as discussed in the following section.

\subsection{Estimated mass, momentum and energy outflow rates}

The distances and sizes of the individual UFO components (or clumps) were constrained in \S4.3, enabling estimates of the mass outflow rate ($\dot{M}_{\rm out}$), momentum outflow rate ($\dot{P}_{\rm out}$), and energy outflow rate (or kinematic power; $\dot{E}_{\rm K}$).
The mass outflow rate is calculated as
\begin{equation}
\dot{M}_{\rm out} = 4 \pi \mu m_{\rm p} R^2 n F_{\rm cov} F_{\rm vol} v_{\rm out}~~~, 
\end{equation}
where $\mu$ is the mean atomic mass per proton, fixed at 1.4, $m_{\rm p}$ is the proton mass, and $n$ is the number density calculated as $n = N_{\rm H}/\Delta R$.
We adopt a covering fraction of $F_{\rm cov} = 1$, based on the fit with the XSTAR emission model for the broad emission line at $\sim$7.4~keV as shown in \S3.1.
The volume filling factor, $F_{\rm vol}$, can be approximated as $F_{\rm vol} \sim N (\Delta R/2)^3 / R^3$, where $N$ is the total number of clumps.
Following the discussion in \cite{XRISM2025}, if we define the multiplicity as $M = N \pi (\Delta R/2)^2/(4\pi R^2)$, the volume filling factor can be rewritten as  $F_{\rm vol} = 4 M (\Delta R/2)/R$.
Assuming $M = 1$ for each zone, the volume filling factors are calculated to be $F_{\rm vol,1} = 0.04$, $F_{\rm vol,2} = 0.14$, and $F_{\rm vol,3} = 0.14$ for Zones~1, 2, and 3, respectively.
The mass outflow rates for Zones 1, 2, and 3 are estimated as $\dot{M}_{\rm out, 1} \sim  1.6~M_{\odot}$~yr$^{-1}$, $\dot{M}_{\rm out, 2} \sim 10.0~M_{\odot}$~yr$^{-1}$, and $\dot{M}_{\rm out, 3} \sim 8.4~M_{\odot}$~yr$^{-1}$, respectively. 
Thus, the total mass outflow rate of the UFOs is estimated to be $\dot{M}_{\rm out} \sim 20.0~M_{\odot}$~yr$^{-1}$.
In comparison, the mass outflow rate of the CO molecular outflow in the host galaxy has been reported to be $\dot{M}_{\rm out, mol} \sim 219~M_{\odot}$~yr$^{-1}$ (\citealt{Fluetsch2019}; \citealt{Smith2019}), indicating that the mass carried by the UFO-driven outflow is substantially lower than that of the molecular outflow.

The momentum outflow and energy outflow rates are calculated as
\begin{equation}
\dot{P}_{\rm out} = \dot{M}_{\rm out} v_{\rm out}~~~, 
\end{equation}
and
\begin{equation}
\dot{E}_{\rm K} = \frac{1}{2} \dot{M}_{\rm out} v^2_{\rm out}~~~, 
\end{equation}
respectively. 
Therefore, the momentum outflow rates for Zones 1, 2, and 3 are estimated as $\dot{P}_{\rm out, 1} \sim  2.4 \times 10^{35}$~dyn, $\dot{P}_{\rm out, 2} \sim  1.9 \times 10^{36}$~dyn, and $\dot{P}_{\rm out, 3} \sim  2.3 \times 10^{36}$~dyn, respectively, giving the total momentum outflow rate of $\dot{P}_{\rm out} \sim  4.4 \times 10^{36}$~dyn. 
On the other hand, the kinematic powers for Zones 1, 2, and 3 are estimated as $\dot{E}_{\rm K,1} = 2.7 \times 10^{44}$~erg~s$^{-1}$, $\dot{E}_{\rm K,2} = 2.9 \times 10^{45}$~erg~s$^{-1}$, and $\dot{E}_{\rm K,3} = 4.9 \times 10^{45}$~erg~s$^{-1}$, respectively. 
The total kinematic power is therefore $\dot{E}_{\rm K} = 8.0 \times 10^{45}$~erg~s$^{-1}$, which is comparable to $L_{\rm Edd}$ for $M_{\rm BH} = 2.5 \times 10^7~M_{\odot}$. 
According to \cite{Fluetsch2019} and \cite{Smith2019}, the momentum outflow rate and kinematic power carried by the CO  molecular outflow are  $\dot{P}_{\rm out, mol} \sim  7 \times 10^{35}$~dyn and $\dot{E}_{\rm K, mol} = 2 \times 10^{43}$~erg~s$^{-1}$, respectively. 
Therefore, the estimated momentum outflow rate of the UFOs exceeds that of galactic molecular outflows by an order of magnitude, while the energy outflow rate exceeds it by a few orders of magnitude. 
As discussed in \S4.2, the present observation shows a stronger presence of UFO components compared to previous observations, indicating that UFO activity exhibits significant temporal variability. 
The intermittent nature of UFO activity could thus be a potential cause of their higher momentum and energy outflow rates relative to those of galactic molecular outflows.
In this context, considering momentum-driven AGN feedback, the duty cycle is estimated to be around 15\%. 
Another possibility is that the energy transfer from the UFO components to the galactic molecular outflow is inefficient. 
Given that similar results have been obtained for the quasar PDS~456 \citep{XRISM2025}, it is expected that powerful UFOs are already formed during the ULIRG phase, and that AGN feedback, likely sustained by these UFOs, suppresses star formation in the host galaxy, thereby driving its evolution into a quasar. 

\section{Conclusions}

We observed the nearly type-2 AGN in the ULIRG IRAS~05189$-$2524 with XRISM and successfully detected three powerful UFO components with distinct bulk velocities and velocity dispersions, thanks to the high energy resolution of the onboard X-ray microcalorimeter.
Specifically, we identified Zone~1 with $v_{\rm bulk} \sim 0.076c$ and $\sigma_v \sim 800$ km s$^{-1}$, Zone~2 with $v_{\rm bulk} \sim 0.101c$ and $\sigma_v \sim 4200$ km s$^{-1}$, and Zone~3 with $v_{\rm bulk} \sim 0.143c$ and $\sigma_v \sim 5800$ km s$^{-1}$.
By dividing the data into the first and latter intervals to investigate UFO variability, we identified possible variability in the UFO parameters, with changes detected in $v_{\rm bulk}$, $\sigma_v$, and $\log \xi$ of Zone~3.
Simultaneous observations with the X-ray CCD camera revealed a relatively soft broadband continuum with a photon index of $\Gamma \sim 2.3$, along with dramatic variability on timescales of $\sim 10$~ksec occurring multiple times at intervals of $\sim40$~ksec.
These characteristics are consistent with accretion taking place near the Eddington limit. 
ULIRGs like IRAS~05189$-$2524 are thought to represent a short-lived evolutionary stage following gas-rich galaxy mergers, during which intense starburst and quasar activity co-exist before star formation is quenched by AGN feedback.
Our findings indicate that powerful, multi-velocity UFOs are launched by near-Eddington accretion during this critical transitional phase.

We constrained the structures of the UFOs based on their multiple bulk velocities and velocity dispersions, and estimated that all three components are located at less than $\sim 1800~R_{\rm S}$, with spatial extents of $\sim 120~R_{\rm S}$ or less.
Using these constraints, we estimated the mass, momentum, and energy outflow rates by the UFOs.
Notably, the energy outflow rate is estimated to be $\dot{E}_{\rm K} \sim 8.0 \times 10^{45}$~erg~s$^{-1}$, which is comparable to the Eddington luminosity and exceeds that of the large-scale molecular outflow in the host galaxy by a few orders of magnitude. Furthermore, the momentum outflow rate, $\dot{P}{\rm out} \sim 4.4 \times 10^{36}$ dyn, is also an order of magnitude higher than that of the galactic molecular outflow. 
As a result, it has been revealed that such powerful UFOs are already well developed during the ULIRG phase. Given that similar results have been obtained for the quasar PDS~456, it is suggested that these strong UFOs are sustained throughout the evolutionary process, during which AGN feedback suppresses star formation in the host galaxy and drives its transition into a quasar.

Since additional UFO components beyond the three identified zones may exist and temporal variability of the UFO components is also expected, future observations of IRAS~05189$-$2524 with XRISM, as well as next-generation X-ray microcalorimeter missions such as NewAthena \citep{Cruise2025}, will be essential for advancing our understanding of UFOs in starburst merging galaxies.

\section*{Acknowledgments}
The authors sincerely thank Dr. James Reeves for serving as the referee and for providing valuable comments and suggestions that greatly improved this paper.
The authors are grateful to Dr. Chris Done for reviewing the manuscript as part of the internal review process of the XRISM science team.
The authors thank Dr. Kazumi Kashiyama, Masahiro Takada, Kunihito Ioka, Masaomi Tanaka, Taiki Kawamuro, Takeo Minezaki, Hiroaki Sameshima, Mitsuru Kokubo, Takashi Horiuchi, and Shoichiro Mizukoshi for valuable discussions.
This study is supported by Japan Society for the Promotion of Science (JSPS) KAKENHI with Grant numbers 19K21884, 20H01947, 20H01941, 21K13958, 23K20239, 20KK0071, 24K00672, 25H00660, and Yamada Science Foundation (MM). 
The material is based upon work supported by NASA under award number 80GSFC24M0006.

\bibliography{iras05189}

\end{document}